\documentstyle[12pt,epsf]{article}

\def\bp{{\bf p}}
\def\br{{\bf r}}

\def\bx{{\bf x}}

\def\bE{{\bf E}}

\newcommand{\be}{\begin{equation}}
\newcommand{\ee}{\end{equation}}
\newcommand{\bea}{\begin{eqnarray}}
\newcommand{\eea}{\end{eqnarray}}

\newcommand{\p}{\partial}
\begin{document}
 
\vskip 2cm
\begin{center}
\Large
{\bf  Nonperturbative QCD Vacuum Effects } \\
{\bf in  Nonlocal Quark Dynamics } \\
\vskip 0.5cm
\large
 Ken Williams \\
{\tiny \em General Delivery/  Ojai, CA. 93023 } \\
\end{center}
\thispagestyle{empty}
\vskip 0.7cm
 
\begin{abstract}
 
  A straightforward calculation reveals the essentially nonlocal
  character of the leading heavy $ Q\bar{Q} $ interaction arising from
  nonperturbative gluon field correlations in the model of a
  fluctuating QCD vacuum. In light of this quarkonium spin splitting
  ratio predictions which have supported the scalar confinement ansatz
  are reconsidered as a specific example of possible consequences for
  spectroscopy.

\end{abstract}
 
\newpage

\section{Nonlocal Dynamics}

A great deal of work has gone into the effort to describe the spectrum
and high energy scattering of hadrons as bound states of quarks and
gluons.  It remains an open and interesting problem. Part of the
trouble seems to lie in the historical identification of particle
field theory with its familiar perturbation expansion, while
nonperturbative phenomena such as confinement are thought to play an
important role in the QCD bound state.  The controlling aim to obtain
an understanding in a formulation of the fundamental Quantum Field
Theory has led to the development of nonperturbative phenomenology
reflecting the effective degrees of freedom for a given arrangement, 
where those degrees less relevant are frozen or integrated out of the
theory.  Unfortunately there is no unique well-defined program by
which this is done. Two competing points of view among many are given
by potential models on one hand and the method of sum rules on the
other. The former assumes a local interaction in terms of quark
coordinates, while the gluon field figures in more directly in the
later.

It was shown some time ago by Voloshin and Leutwyler separately
\cite{vl} that large scale fluctuations of the QCD vacuum cannot be
described adequately by local potentials.  The conclusion found
support in the subsequent leading relativistic interaction Hamiltonian
of Eichten and Feinberg \cite{ef} (see also Lemma in \cite{g}) and
refinement at the hands of Marquard and Dosch \cite{md} who considered
two extreme cases for the vacuum field's correlation length relative
to those of the heavy quarks bound in a meson. Roughly, for $ T_g <<
T_q $ a local description was found to be justified, though for $ T_g
>> T_q $ a nonlocality appears making the sum rule approach more
suitable.

Potential model builders have taken this to validate the use of local
potentials for sufficiently heavy quarks, which is quite right. A
quantitative sense of the validity in terms familiar to the language
of potential models might be obtained by considering the interaction
as a double expansion in the ratios of the two time scales with a
common scale, say $ \Lambda_{QCD} $, appropriate for the above limits.
To any specified order in $ T_q $ then the limits now read, $ T_g \to
0 $ and $ T_g \to \infty $ , for local and nonlocal couplings
respectively. Finite $ T_g \approx 1.0 $ $ GeV^{-1} $ ( from the
lattice \cite{lat}) then requires a generalization of the analysis in
line with intermediate correlations.

This is easily carried out beginning from the expression given in
\cite{md} (not to be re-derived here ) for the Schwinger function of a
singlet $ Q\bar{Q} $ pair in an external color field. In the leading
dipole approximation \cite{vk}
\bea
 G &\sim & \int [ d\bx ]\exp \left[-\frac{g^2_s}{36}\int^t_0
  \int^t_0 \bx(t_1)\cdot\bx(t_2) \langle \bE^a(t_1)\cdot\bE^a(t_2)
  \rangle_E dt_1 dt_2 \right] \\ & \approx & \int [d\bx ] \exp
\left[\int^t_0 dt_1 (\bx^2 - 2\bx\cdot\frac{\bp}{m} \zeta ) \right]
\label{green} 
\eea
with
\bea
\zeta &=& - \frac{g^2_s}{36} \int^\infty_0 d\tau \tau \langle
\bE^a(t_1)\cdot\bE^a(t_2) \rangle_E \quad , \qquad 
\quad \tau \equiv t_1 - t_2
\eea
assuming the correlator falls off rapidly for large Euclidean time
differences.  The integral over $ -\frac{g^2_s}{36} \langle
\bE^a(t_1)\cdot \bE^a(t_2) \rangle_E $ has been normalized to 1 and $
\bx(t_2) \approx \bx(t_1) - \tau \dot{\bx} (t_1) $ to lowest order in
quark velocity has been used.

Expression (\ref{green}) is not difficult to interpret. The
nonlocality enters scaled by the time rate of vacuum correlations. For
example, should the fields correlate adiabatically with respect to
quark motion, corresponding to a stochastic delta correlation, or
white noise, so that $ \frac{\bp}{m} \zeta \approx 0 $, a local
potential emerges. But only in this limit. {\em All } other
correlations lead to nonlocal dynamics the degree of which measured by
gluon degrees of freedom as they occur in the fluctuating vacuum. For
a general discussion of the effect in the context of the flux tube
picture see Isgur \cite{isgur}.

A convenient parameterization for the statistical distribution is
provided by the Stochastic Vacuum Model \cite{sim}
\bea
\langle \bE^a(t_1) \cdot \bE^a(t_2) \rangle_E  = & 3 \beta [
D(\tau ) + D_1(\tau ) + \tau^2 \frac{\p D_1 }{ \p \tau^2 } ]
\eea
with
\bea
D, D_1 \sim \exp( - | \tau | / T_g )
\eea
The model calculation estimates nonlocal contributions to
(\ref{green}) (for $ \frac{ \langle p \rangle }{ m } \approx 0.17 $
) at $ 50 \% $.

\section{ Quarkonium Spin Splittings}

The qualitative success of the Nonrelativistic Potential Model with
linear confinement does not entirely carry over when introduced into
relativistic kinematics. The problem has long been thought to be
related to a neglect of gluon field momentum. Examples in the study of
Regge trajectories are found in \cite{go}. Nonlocal effects might
also be relevant to the question of possible Lorentz structures for
confinement. Evaluation of heavy $ Q\bar{Q} $ spin splittings for a
local potential leads to a widely known argument in favor of dominant
scalar confinement. The quantity of interest is the ratio of $ \chi $
- state masses \cite{sch}
\bea
r &\equiv & \frac{ M_2 - M_1 }{ M_1 - M_0 } \approx 0.5 (exp.)
\eea
with expected values of less than 0.8 or greater than 2.0 for linear
scalar coupling and ranging from 0.8 to 1.4 in the vector case. The
other structures are ruled out due to their incompatibility with
appropriate nonrelativistic limits arising from pure Lorentz sources.
Here the analysis is reconsidered taking nonlocal effects into
account.

In the conventional treatment \cite{gl} one begins with an
expansion of an arbitrary interaction kernel over the five Lorentz
invariant amplitudes ( scalar, pseudoscalar, vector, axialvector and
tensor )in the $ q \otimes \bar{q} $ Dirac space
\bea
 K &=& V_s 1\otimes 1 +V_{ps}\gamma^5 \otimes \gamma_5 + V_v
\gamma^\mu \otimes \gamma_\mu \label{k} \\ && \qquad \qquad \qquad + V_{av}\gamma^\mu \gamma^5
\otimes \gamma_\mu \gamma_5 + \frac{1}{2} V_t \sigma^{\mu \nu }
\otimes \sigma_{\mu \nu } \, . \nonumber
\eea
A $ O(m^{-2}) $ potential is then derived in the usual way by a
Fourier transform of the transformation matrix or a reduction of the
Bethe-Salpeter equation. Both routes however require the use of free
quark propagators of whose adequacy there is some question
particularly in the present nonperturbative \cite{np} nonlocal
\cite{nl} context.  An alternative approach possibly more compatible
with the presence of the external vacuum field might be a nonlocal
minimal coupling, e.g., by way of the instantaneous Salpeter equation.
The question will not be gone into here. For comparison with the local
result and for the sake of simplicity it is useful to proceed in the
usual way as discussed above.  The lowest order nonlocal contribution
to the amplitudes of (\ref{k}) in the center of momentum is
parameterized as
\bea
V_{nl} & = &\{ \Theta_{ij}(\br), p_i p_j\}_{symm}
\eea
with
\bea
\Theta_{ij}(\br) &=& \theta_1(r)\delta_{ij} + \theta_2(r) \hat{r}_i\hat{r}_j
\eea
where $ \br $ is the relative coordinate.  A similar parameterization
is used in \cite{g1} where the nonlocality is proposed to resolve the
outstanding small baryon splitting puzzle. After adding on the Cornell
potential, $ -\frac{4}{3}\frac{\alpha_s}{r} + a r $ , and taking the
simplifying, $ \theta_1 = - \theta_2 \equiv \theta $ , a
straightforward but lengthy calculation yields
\bea
 r_{scalar} &=& \frac{1}{5} \frac{ 8\alpha_s \langle r^{-3}
  \rangle -\frac{5}{2} a \langle r^{-1} \rangle + 5 \langle R_1
  \rangle }{ 2\alpha_s \langle r^{-3} \rangle -\frac{1}{4} \ a \langle
  r^{-1} \rangle + \frac{1}{2} \langle R_1 \rangle } \\
 r_{vector} &=& \frac{1}{5} \frac{ 8\alpha_s \langle r^{-3}
  \rangle + 7 a \langle r^{-1} \rangle + \frac{5}{2} \langle {\cal R }_{(2-1)}
  \rangle }{ 2\alpha_s \langle r^{-3} \rangle + \ a \langle
  r^{-1} \rangle + \frac{1}{2} \langle {\cal R }_{(1-0)} \rangle } 
\eea
with $ R $ and $ {\cal R } $ given in the appendix. It is clearly not
possible to establish a numerical range for these expressions without
more information on the nonlocality.  As it stands the analysis is
indeterminate, favoring no Lorentz structure for confinement over
another.

\section{Summary and Discussion}

The main point emphasized here has been that nonperturbative gluon
degrees generally arise in the hadronic bound state in the form of
nonlocal quark dynamics. This has been shown to follow from a simple
variation on the analysis of \cite{md}.  A corollary is that the
nonlocality appears whenever quark motion is taken into account - e.g.,
in the leading relativistic contributions to the static local
interaction potential.  Consequences for quarkonium spin splitting
ratios have been considered with a nonlocal parameterization.

It should be added for completeness that that the pseudoscalar,
axialvector, and tensor Lorentz structures fail to reduce individually
to suitable static limits does not exclude them apriori from
participation at higher orders.  On the contrary. The nonrelativistic
limit suggests nothing beyond what might be said of itself.  This
observation included leaves the local analysis of \cite{gl}
indeterminate also. While scalar confinement may or may not be the
most simple ansatz ( according to one's sense of the simple ) it is
certainly not an unavoidable conclusion; vector + pseudoscalar, vector
+ axialvector + tensor among other combination spin structures for a
confining {\em local } interaction presumably yield equal or better
predictions.

What is needed of course is an understanding of the mechanism by which
nonperturbative via nonlocal forces enter into the QCD bound state and
effective means by which reliable estimates of the effect can be
obtained. The Wilson loop occurs naturally in gauge invariant
formulations of the state.  It (and so interactions derived from it)
is manifestly nonlocal for nonstatic quarks. Its evaluation in the
Minimal Area Law, the Stochastic Vacuum Model, and Dual QCD have
recently been carried out by Brambilla and Vario \cite{bv}. These are
leading order approximations from three mutually distinct expansions
of the Wilson loop - no one contained entirely within another.

Minimal substitution of the QCD inspired relativistic flux tube
\cite{lo} into the linear Dirac equation \cite{ow} is an example of a
promising nonlocal model with many attractive features: correct Regge
structure and spin orbit sign, to name two. As a model however it is
to be measured against both observation and the fundamental theory. On
the other hand, differences between evaluations of the Wilson loop in
any one of the above mentioned approximations \cite{w} is amenable to
unambiguous, mathematical resolution. These points seem to have been
missed by ref\cite{k}.

\vspace{10mm}
 
\hspace{-8mm} {\sl Acknowledgment}: The author is grateful to
the Rain Community Internet Center of Santa Barbara, California for
the generous access services they provide free to the public.

\section{Appendix}

\bea
{\cal R }_{(2-1)} &=& -6 R_1 -\frac{1}{5} R_2 -\frac{1}{10} R_3 ( 2\frac{\p}{\p r } +\frac{3}{r} ) + \frac{1}{10} R_4 (\frac{1}{r} \frac{\p}{\p r} -\frac{4}{r^2} ) \\
{\cal R }_{(1-0)} &=& -3 R_1 +\frac{1}{2} R_2 +\frac{1}{4} R_3 ( 2\frac{\p}{\p r\
 } +\frac{3}{r} ) - \frac{1}{4} R_4 (\frac{1}{r} \frac{\p}{\p r} -\frac{4}{r^2} \
)
\eea
with
\bea
R_1 &=& \frac{1}{ r^2} \left[ \theta (\frac{\p^2 }{\p r^2 } + \frac{2}{3} \frac{1}{r^2} ) +\theta^\prime ( \frac{\p}{\p r} -\frac{1}{r} ) -\frac{1}{3} \theta^{\prime \prime } \right] \\
R_2 &=& \frac{2}{r} \left[ \frac{1}{r} \theta (3  \frac{\p^2}{\p r^2} +\frac{5}{r}  \frac{\p}{\p r} + \frac{8}{3} \frac{1}{r^2} ) - 2\theta^\prime (  \frac{\p^2}{\p r^2} +\frac{1}{r}  \frac{\p}{\p r} - \frac{2}{3} \frac{1}{r^2} ) + \frac{1}{3} \frac{1}{r} \theta^{\prime \prime } + \frac{1}{3} \theta^{\prime \prime \prime} \right] \\
R_3 &=& \frac{4}{r} \left[ - \frac{1}{r} \theta ( \frac{\p}{\p r} +\frac{3}{2} \frac{1}{r} ) + \theta^\prime  \frac{\p}{\p r} \right] \\
R_4 &=& - \frac{2}{r^2} \theta  
\eea

\end{document}